\documentclass[aps,prl,twocolumn,showpacs,superscriptaddress]{revtex4-1}

\usepackage{amsmath} 
\usepackage{amsthm} 
\usepackage{amssymb}	
\usepackage{graphicx} 
\usepackage[table]{xcolor}
\usepackage{comment}

\def\beq{\begin{equation}}
\def\eeq{\end{equation}}
\def\bea{\begin{eqnarray}}
\def\eea{\end{eqnarray}}

\newcommand{\ket}[1]{\left| #1 \right>} 

\begin{document}

\title{Quantum Control of  Many-body Localized States}
\author{S. Choi}
\affiliation{Physics Department, Harvard University, Cambridge, Massachusetts 02138, USA}

\author{N. Y. Yao}
\affiliation{Physics Department,  University of California Berkeley, Berkeley, California 94720, USA}

\author{S. Gopalakrishnan}
\affiliation{Physics Department, Harvard University, Cambridge, Massachusetts 02138, USA}

\author{M. D. Lukin}
\affiliation{Physics Department, Harvard University, Cambridge, Massachusetts 02138, USA}

\date{\today}

\begin{abstract}
We propose and analyze a new  approach to the coherent control and manipulation of quantum degrees of freedom in disordered, interacting systems in the many-body localized phase. 
Our approach leverages a number of unique features of many-body localization: a lack of thermalization, a locally gapped spectrum, and slow dephasing. 
Using the technique of quantum phase estimation, we demonstrate a protocol that enables the local preparation of a many-body system into an effective eigenstate. 
This leads to the ability to encode  information and control interactions without  full microscopic knowledge of the underlying Hamiltonian. 
Finally, we analyze the effects of weak coupling to an external bath and provide an estimate for the fidelity of our protocol.
\end{abstract}


\maketitle

The coherent control and manipulation of a complex quantum system is one of the central challenges of modern physics. 
The majority of ongoing research focuses on building up this complexity starting from individual, isolated qubits \cite{Ladd:2010kq,Nielsen:2010vu,OBrien:2007wl,Haffner2008155,Jones201191,PhysRevA.57.120,Yao:1jn}.
In contrast, the opposite approach, where one seeks the  coherent manipulation of a strongly interacting system, especially subject to disorder, is generally thought to be intractable. 
The main difficulty is the exponentially growing Hilbert space with a typical many-body eigenstate strongly coupled to a dense set of other states. 
Even when the structure of relevant eigenstates is known, their many-body character generally makes them difficult to manipulate using most experimental controls. 

In this Letter, we explore an alternate approach toward the coherent control of many-body systems. Our approach follows a new paradigm introduced by recent studies of many-body localization (MBL) \cite{Anderson58,
Basko:2006hh,
PhysRevB.75.155111,
PhysRevB.77.064426,
Huse_2010,
PhysRevB.81.134202,
Bardarson:2012gc,
abanin_2013_sep,
abanin_2013,
PhysRevB.87.134202,
Bahri:2013ux,
nayak2013,
Huse_2013,
Sarang1408,
PhysRevB.90.174302,
yao2014quasi,
Chandran:2014wu,
Nandkishore:2014vg,
Pino:2014bn,
Khemani:2014wz,
PhysRevX.4.041021,
burin2006energy,
burin2015many,
Ros2015420,
PhysRevLett.114.217201,
2015arXiv150103501P}.
In the presence of strong disorder, the many-body eigenstates of an isolated, interacting system can be localized in Fock space \cite{PhysRevB.81.134202}.
These MBL eigenstates  exhibit a discrete local spectral response,  suggesting that one can coherently manipulate and store quantum information (using external controls with both finite spatial and spectral resolution) \cite{Sarang1408,Chandran:2014wu,PhysRevB.90.174302}. 
This  is remarkable because it implies that, in the presence of sufficiently strong disorder, local quantum bits   naturally emerge from a strongly interacting many-body system \cite{Nandkishore:2014vg,PhysRevB.90.174302}.

Further, in the MBL phase, these  emergent qubits interact with one another via diagonal (i.e., dephasing) interactions; for two qubits separated by a distance $d$, this interaction time scales as $\sim \tau \exp(d/\xi)$, where $\tau$ is a characteristic timescale and $\xi$ is the localization length \cite{Huse_2013,abanin_2013}. In this respect, the MBL phase differs crucially from, e.g., a conventional Anderson insulator, in which such interactions are absent \cite{Anderson58,Bardarson:2012gc,Knap2014}. 
This weak interaction brings about two implications for quantum control: (i) the coherence time of a single MBL qubit can be extremely long and (ii) dephasing can be leveraged as a way to  coherently couple  spatially separated MBL qubits
\cite{abanin_2013}. 
Unfortunately, in general,  the parametric dependence of both the  coherence and  dephasing times are identical.
 To make use  of many-body localization as a platform for quantum control,  one must devise a method to implement interactions (e.g. two-qubit quantum gates) on much shorter timescales than the decoherence.

\begin{figure}[t]
\includegraphics[width=3.3in]{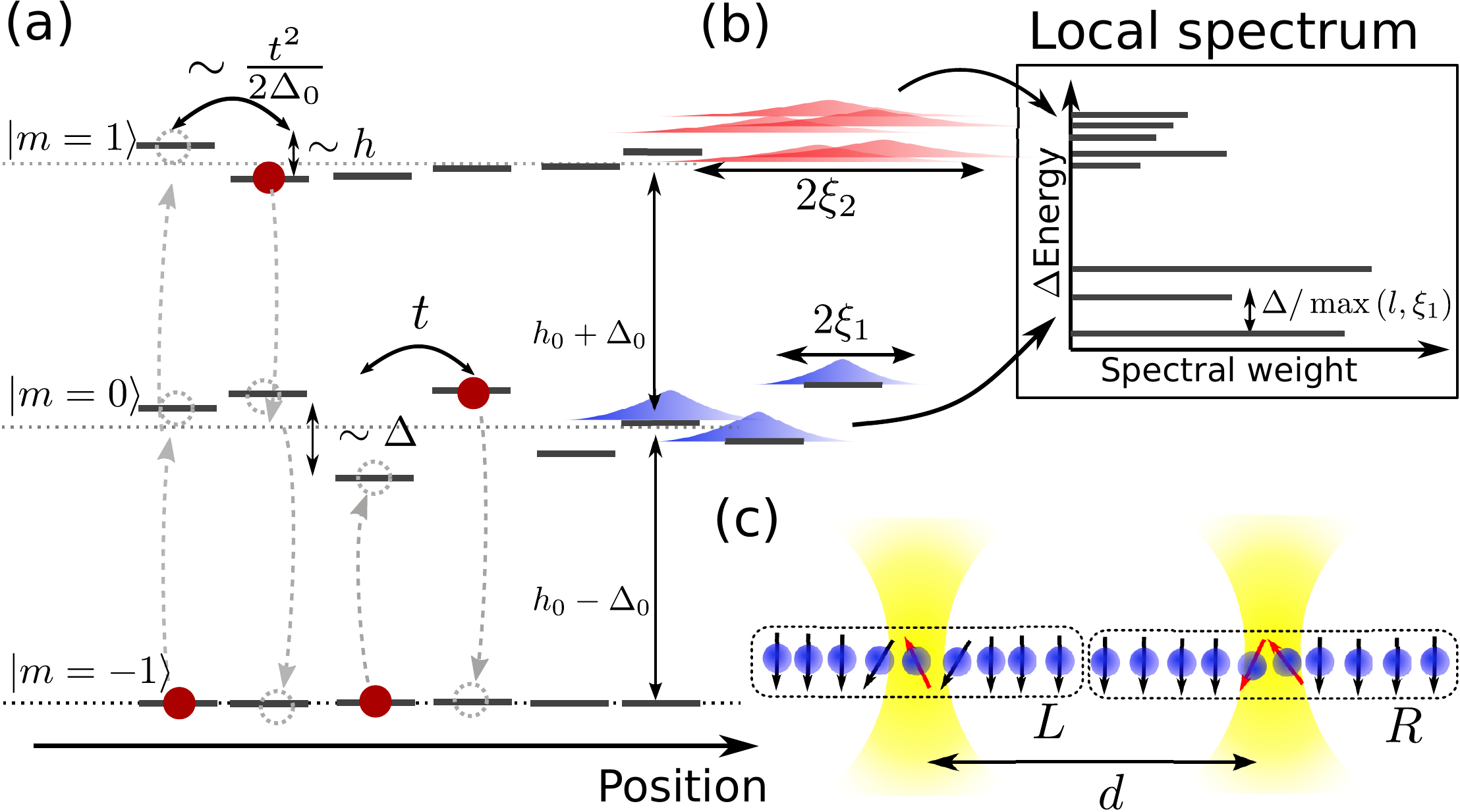}
\caption{(a) Schematic energy spectrum of the disordered spin-1 model. (b) A discrete local spectrum with  excitations exhibiting different localization lengths. (c) External electromagnetic driving (yellow beam) defines the location of an MBL qubit.}
\label{fig:level_diagram}
\end{figure}

Our approach relies upon three key ingredients. 
First, the local spectral gap of the MBL phase enables one to encode and manipulate quantum information using controls of  both finite spatial and spectral resolution. 
Second, we demonstrate the ability to enhance interactions between MBL qubits by exploiting variations in the localization length of different eigenstates. 
Finally, we utilize quantum phase estimation and  projective measurements to locally prepare the system in a many-body eigenstate ensuring that the interactions between spatially separated MBL qubits is coherent. In contrast to previous work \cite{PhysRevX.4.041021}, our approach uses the intrinsic dephasing interactions of the MBL phase as a resource to purify the system.
The combination of these ingredients can in principle, form the basis for a many-body localized quantum information processor. We find that such an approach can  achieve a fidelity that asymptotically approaches the conventional method using isolated qubits.  Perhaps more significantly, the local encoding of information provides new routes for studying and quantifying dynamics in the MBL phase. For example, one can imagine directly probing the decay of both the classical and quantum information, extracting the characteristic decay length for each \cite{MariCarmen15}, thereby gaining new insights into the nature of MBL systems.  

\emph{Spin-1 model}---To illustrate the key idea of our approach, we start with a simple model Hamiltonian that clearly exhibits two separate localization lengths. 
We will utilize excitations with a short localization length (``qubit'' excitation) to encode  information; meanwhile, to implement gates, we envision off-resonantly coupling such qubit excitations to a more delocalized excitation (``bus'' excitation), thereby enhancing interactions on demand \cite{comment:ion_trap,Haffner2008155}.

Consider a chain of spin-1 particles with Hamiltonian,
\begin{equation}
H = \sum_{i=1}^{N} h_i S_i^z + \Delta_i \left(S_i^z\right)^2 + 2t \vec{S}_i \cdot \vec{S}_{i+1}\label{eqn:hamiltonian}
\end{equation}
where $N$ is the total number of sites (Fig.~1a). Disorder exists in both the magnetic field strength $h_i$ and the single spin anisotropy $\Delta_i$, which are randomly drawn from uniform distributions $[h_0 -h, h_0 + h]$ and $[\Delta_0-\Delta, \Delta_0 + \Delta ]$, respectively.
As the Hamiltonian conserves total $z$-magnetization, $\sum\nolimits_i S_\textrm{i}^z$, we can  define a logical ground state $\ket{G}$ as the eigenstate with all spins in $\ket{S_i^z=-1}$.
We focus on the regime in which the field disorder is much weaker than the anisotropy disorder, with a hierarchy, 
\beq\label{scales}
t \ll h \Delta_0/\Delta \ll \Delta \ll \Delta_0 \alt h_0.
\eeq

In this limit, there are two types of well-defined single-particle excitations. The first (qubit-type) consists of a single spin flipped into the $\ket{S_i^z = 0}$ state.  This  excitation moves in a disorder potential of strength $\Delta$,  analogous to Anderson localization \cite{Anderson58}.
 As such, all single-particle states are localized and in the limit $t \ll \Delta$, the typical localization length of an excitation is given by $\xi_1^{-1} \sim \log{\Delta/t}\label{eqn:loclength1}$. 
The second single-particle excitation (bus-type) consists of a single spin flipped into the $\ket{S_i^z = +1}$ state; it can be regarded as a bound state of two qubit-type excitations, with binding energy $\sim \Delta_0$ (Fig.~1a). 
A bus excitation propagates  via a second-order hopping process of amplitude $\sim t^2/(2\Delta_0)$. It, too, can be regarded as a localized single-particle excitation, with localization length,
$\xi_2^{-1} \sim \log{(h \Delta_0/t^2)} = \xi_1^{-1} - \log{[t \Delta/(h \Delta_0)]}.\label{eqn:loclength2}$ %
Under the hierarchy of equation~\eqref{scales}, $\xi_2 \gg \xi_1$ (Fig.~1b). 

We now turn to the case of multiple excitations.
For sufficiently small $t$ or sufficiently low energy densities, the many-body eigenstates of this system are expected to exhibit many-body localization. The effective interaction between two excitations in the MBL phase falls off exponentially with distance; in the limit of low energy density, it is simply given by the overlap of their wavefunctions.
Thus, for two qubit-type excitations a distance $d$ from one another, the interaction strength is
\beq\label{eqn:decoherenceTime}
\delta_I \sim t \exp(-d/\xi_1),
\eeq
while for two  bus-type excitations, the interactions scale as  $\sim t \exp(-d/\xi_2)$.

\emph{Single qubit encoding and manipulation}---In order to encode $N_q$ qubits, we partition a given disordered system of size $L$ into $N_q = L/d$ smaller segments of length $d > \xi_2 \gg \xi_1$. We select a single qubit-type excitation centered near the middle of each segment, and use it to encode the state of the MBL qubit (Fig.~1c).
An arbitrary superposition of qubit states will dephase on a time scale $1/\delta_I$, and all gate operations must be fast compared to this timescale~\cite{Huse_2013,abanin_2013}.

The coherent manipulation of a single qubit state is feasible with local external control of  finite spatial $\ell$ and spectral $\delta \omega$ resolution \cite{Chandran:2014wu}.
After initializing the system to the logical ground state $\ket{G}$, one applies a time-dependent electromagnetic field of frequency $\omega$ (Fig.~\ref{fig:level_diagram}c); to within exponentially small corrections, this transverse field couples $\ket{G}$  with only a finite number $\sim \textrm{max}(\ell, \xi_1)$ of localized eigenstates,  as depicted in Fig.~\ref{fig:level_diagram}b.

Thus, one can induce a high-fidelity  transition between $\ket{G}$ and a specific eigenstate by weakly driving the system with $\omega$ tuned to the particular many-body  transition $\omega_L$. The requirements for the spatial and spectral resolution are: (i) $\ell < d$ and (ii) $\delta \omega < \Delta / \textrm{max}(\ell,\xi_1)$.
In this encoding scheme, information readout is performed by directly measuring the local spin polarization. In the limit $\xi_1 \rightarrow a$, this corresponds to the measurement of a single spin. When $\xi_1 > a$, the total polarization in the region determines the qubit state.

\emph{Two-qubit gates}---To perform universal quantum control, we only need to demonstrate a controlled phase gate. As previously mentioned, such a gate can in principle,  be achieved via the dephasing interactions of Eq.~\eqref{eqn:decoherenceTime}.
However, since decoherence occurs at the same rate, one must effectively enhance this  interaction strength.
Since the dephasing between bus excitations occurs at a much faster rate $te^{-d/\xi_2}$,  the effective interaction strength  can be enhanced by several orders of magnitude upon dressing \cite{saffman2010quantum} a qubit-type excitation with a bus-type excitation. 
This enhancement ratio is given by,
\begin{equation}
\frac{ \delta^\textrm{driven}_I}{ \delta_I} \sim  \frac{|\Omega|^2}{|\delta|^2} e^{d \left(\frac{1}{\xi1} - \frac{1}{\xi_2}\right)}\label{eqn:interactionEnhancement}
\end{equation}
where $\Omega$ is the Rabi frequency and $\delta$ is the detuning of the external driving field.
A numerical demonstration of this enhancement in the simplified spin-1 model is provided in the Supplemental Material \cite{suppinfo}.

\emph{Generic MBL system}---We now generalize our protocol from the spin-1 model to any MBL system with a fully localized spectrum \cite{Anderson58,
Basko:2006hh,
PhysRevB.75.155111,
PhysRevB.77.064426,
Huse_2010,
PhysRevB.81.134202,
Bardarson:2012gc,
abanin_2013_sep,
abanin_2013,
PhysRevB.87.134202,
Bahri:2013ux,
nayak2013,
Huse_2013,
Sarang1408,
PhysRevB.90.174302,
yao2014quasi,
Chandran:2014wu,
Nandkishore:2014vg,
Pino:2014bn,
Khemani:2014wz}. Such systems can be represented in terms of conserved local spin-1/2 degrees of freedom termed ``l-bits''~\cite{Huse_2013,abanin_2013}:

\begin{equation}
 H_\textrm{mbl} = \sum_i \bar{h}_i \tau_i^z + \sum_{i,j} \bar{J}_{ij} \tau_i^z \tau_j^z + \sum_{i,j,k} \bar{K}_{ijk} \tau_i^z\tau_j^z\tau_k^z \cdots \label{eqn:huseModel}
\end{equation}
where the $\tau_i$  are spin-1/2 operators describing localized excitations of size $\tilde{\xi}$, $\bar{h}_i$ is the energy of each excitation, $\bar{J}_{ij}$ is the two-body interaction shift, etc. 
An l-bit overlaps with $\sim \tilde{\xi}$ microscopic degrees of freedom, and can therefore be individually addressed (up to exponentially small infidelities) using a drive with spectral resolution greater than $t/2^{\tilde{\xi}}$. In addition, the l-bits have effective diagonal interactions that also fall off exponentially, with a localization length $\xi$ that varies from site to site and (unlike the l-bit size $\tilde{\xi}$) from eigenstate to eigenstate~\cite{Huse_2013}. All  of the qualitative arguments from the spin-1 model extend to the generic case, provided (a)~there are excitations that overlap spatially but have different localization lengths $\xi$, and (b)~one can initially prepare the system in a particular eigenstate.

To understand the first of these issues, we  numerically study  the variation of localization lengths for the Hamiltonian,
\begin{figure}
\includegraphics[width=3.3in]{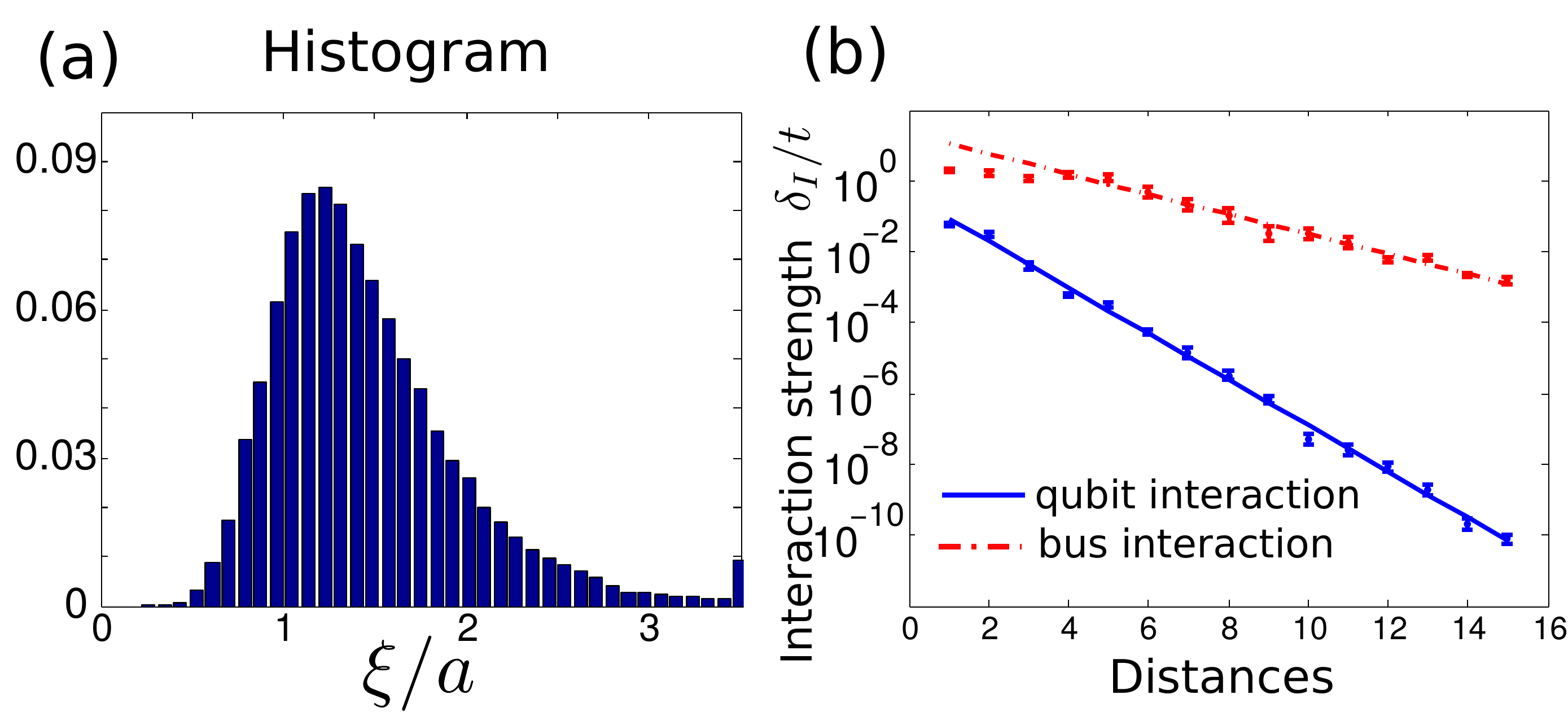}
\caption{Numerical study of localization lengths in the low spin density, weakly localized regime of  Eq.~\eqref{eqn:generic_model}. (a) Histogram of localization lengths in units of lattice spacing. Since the model conserves total $S^z$, one can choose to work in manifolds with a fixed density of excitations. Here, we consider two spin-flips. 
(b) Decay of the interaction strength  as a function of distance for both qubit and bus excitations. }
\label{fig:fig_3}
\end{figure}
 \beq
 H = \sum_i 2 t \vec{S}_i \cdot \vec{S}_{i+1} + h_i S^z_i \label{eqn:generic_model}
 \eeq
where $h_i$ is randomly drawn from a uniform distribution $[-\Delta, \Delta]$. This model is ``generic'' in that we have not engineered its localization lengths to have any particular features.
 In Fig.~\ref{fig:fig_3}a, the histogram of localization lengths $\xi$ is shown for  low excitation density (up to two spin-flips) with weak disorder $\Delta/t= 3$ \cite{suppinfo}. The distribution exhibits heavy tails, evincing the fact that there is significant probability to find $\xi$ longer than its average value.
 It turns out that  variations always exist (even at high excitation densities) and are largest near the localization transition (see \cite{suppinfo} for a quantitative analysis).

We note that our protocol  does not require a parametric separation between long and short localization lengths; rather,  one can optimize $d$ to achieve as large an interaction enhancement as desired.
To quantify the enhancement of the interaction strength, we numerically extract the localization length of typical qubit and bus excitations. We  diagonalize systems up to length $L =50$ with at most two excitations and perform $100$ disorder averages.
For each disorder realization, we apriori choose approximate positions for the left and right MBL qubits (at $L$ and $R$). 
Starting from the polarized state $|G\rangle$, we  identify computational qubits with significant local support around $L,R$. 
These  basis states are chosen as energy eigenstates that have significant overlap with local excitation operators $\Sigma^{L(R)} =  S^x_{L(R)-1} + S^x_{L(R)} + S^x_{L(R)+1} $. In particular, we select states $(\phi^{L}, \phi^{R}, \phi^{LR} )$, such that $|\langle \phi^L | \Sigma^{L} | G \rangle| ^2 , |\langle \phi^R | \Sigma^{R} | G \rangle| ^2 , |\langle \phi^{LR} | \Sigma^{L} \Sigma^{R}  | G \rangle| > 10^{-2}$; moreover, to ensure that the MBL qubits  are self-consistently defined, we require $|\langle \phi^{LR} | \Sigma^L |\phi^R \rangle|, |\langle \phi^{LR} | \Sigma^R |\phi^L \rangle| >10^{-2}$. 

This enables us to directly estimate the effective interaction between the MBL qubits from  energy eigenvalues, with $\delta_I \equiv E_{LR} -E_L - E_R + E_G$. As the qubit(bus)-type excitation, we select basis states, $( \phi^{L}, \phi^{R}, \phi^{LR} )$ that exhibit the smallest(largest) interaction strength \cite{suppinfo}. As depicted in Fig.~\ref{fig:fig_3}b, bus-type excitations interact  significantly more strongly than qubit-type excitations, enabling multiple quantum gates to be performed within a decoherence window. While this simplified simulation does not  account for the dynamical  ``dressing'' of the qubits, the nature of the qualitative enhancement is the same: it comes from  intrinsic variations of the many-body localization length (see \cite{suppinfo} for full numerical simulations). 

\emph{Initial state preparation}---To enable fully coherent evolution, the interactions between  MBL qubits must be coherent. This is indeed the case when the system is prepared in a fiducial many-body eigenstate; however, such a preparation is very difficult for a generic (high temperature) system. 
Since the MBL phase has a simple description in terms of local conserved quantities, each many-body eigenstate can be labeled by specifying the values ($\pm 1$) of all $\tau$ spins.
The question is then: how can one efficiently prepare the system into a desired $\tau$-spin superposition, starting from an arbitrary many-body state, with only local control over small sub-regions $A$ and $B$. 
Let us label the two excitations $L$ and $R$ (in each region) to be our MBL qubits.  The effective interaction between these qubits takes the form $\delta_I= J_{LR} + \sum_{k \notin \{A, B\}} K_{LkR} \tau_k + \ldots$ for each $\tau$-spin configuration \cite{Huse_2013}. 
Thus, starting from a generic many-body state, the effective interaction varies from eigenstate to eigenstate and cannot enable a coherent quantum gate.

Interestingly, the interaction itself can be used as a resource to purify the entire system of $\tau$-spins; in particular, by observing the interaction strength between the MBL qubits, one can effectively perform a quantum non-demolition measurement of the $\tau$-spin configuration.
Such an observation can be done via a modified spin-echo protocol~\cite{Knap2014}, which projects the $\tau$ spins onto a set of configurations that  have the same $\delta_I$, up to the precision of the measurement.

Specifically, using an adaptive phase estimation algorithm \cite{Nielsen:2010vu,PhysRevX.4.041021,PhysRevA.74.032316}, one can repetitively measure the interaction strength, $\delta_I = 2\pi t  \sum_{\alpha=1}^M s_\alpha 2^{-\alpha}$ (in binary),  to a precision set by its smallest significant digit $s_M$. To measure each digit, $s_k$, requires a time, $ T_k = \frac{2^k}{t} $, yielding a total measurement time,
\beq
T_{tot} = \sum_{k=1}^M T_k = 2 \tau_M (1 - 2^{-M}).
\eeq
A few observations are in order: (1) this procedure is extremely efficient since the number of measurements $M$ scales logarithmically with the desired precision and (2) the total measurement time $T_{tot}$ is also the time-scale over which the MBL qubits can now be expected to interact coherently. 
This result implies that one can perform coherent quantum manipulations in the infinite temperature MBL phase with a preparation overhead scaling only linearly in time. 

\emph{Imperfections}---In what follows, we analyze a variety of realistic imperfections and provide a quantitive estimate for achievable fidelities in a number of experimental systems. In particular, we will consider the effect of finite spatial($\ell$) and spectral($\delta \omega$) resolution,  leading to:  (1) imperfect initialization, (2) off-resonant excitation, and (3) population loss into nearby modes.

For single qubit gates, the fidelity is given by, 
\beq
 \mathcal{F}_1 \simeq 1- \frac{\ell}{\xi_1} \left( \frac{1}{T_2 \mathcal{E}} \right)^2,
\eeq
where $T_2$ characterizes the extrinsic decoherence time and $\mathcal{E}$ represents the local spectral gap. The lack of individual addressability manifests as a ratio of the spatial extent to the qubit localization length $\ell/\xi_1$, while off-resonant excitations induce an error $\sim 1/(T_2\mathcal{E})^2$.

Turning to the fidelity for two qubit gates, we note that finite resolutions (spatial and spectral) bring about two consequences, namely, an optimal choice of qubit separation $d_{opt}$ and a renormalized bus localization length $\xi_2 \rightarrow \xi'$ \cite{suppinfo}. The former arises when the decoherence rate begins to dominate the bare interaction strength, while the latter occurs for line-widths larger than the local spectral gap. The physics of this latter case is analogous to  coupling a single bound state (qubit excitation) with a multi-particle continuum (bus excitation) of effective mass $m^* \simeq 1/ta^2$, and leads to a  renormalized excitation size $\xi' \approx a \sqrt{t T_2}$ \cite{Goban:1eq,PhysRevB.43.12772,PhysRevLett.64.2418, suppinfo}. Combining these two effects gives an overall fidelity, 
\beq
\label{two_qubit_fidelity}
\mathcal{F}_2 \sim  1 - (t /\Gamma)^{-1+\xi_1 /\xi_2^{eff}},
\eeq
where $\Gamma \simeq  \ell^2/t T_2^2 a^2  + 1/ T_2$ is an effective decoherence rate estimated from Fermi's Golden rule and $\xi_2^{eff} = \xi_2 (\xi')$ for spectrally (un)-resolved bus excitations.

Interestingly, these  bounds are consistent with traditional quantum information processing schemes based upon isolated qubits. 
In particular, for $\xi_1,\ell \rightarrow a$ and $\xi_2 \rightarrow \infty$, one discovers 
$\mathcal{F}_2^{opt} \simeq 1 - (t T_2)^{-1 + \frac{1}{\sqrt{t T_2}}}$. This corresponds to two tightly localized qubits (e.g. atoms or bound states) interacting via a band of delocalized states  (e.g. phonons or photons).

\emph{Fidelity Estimates}---Our protocol applies most readily to quantum optical systems with local addressing. A number of such platforms are promising candidates for realizing many-body localization, including ultracold atoms, dipolar molecules, superconducting qubits, and solid-state spins \cite{Yao:2013vc,PhysRevB.76.052203,Yan:2013fn,Ni:2008dk}. 
In the case of ultracold atoms, a direct implementation of a spin system is feasible via multi-component Fermi- or Bose-Hubbard models. From recent experiments \cite{PhysRevLett.104.080401,Trotzky:2008jy,Fukuhara:2013hq}, the spatial resolution $\ell \sim a$, the typical superexchange interaction strength $t\sim 10$Hz, and the coherence limited by particle loss $T_2 \sim 3$s are feasible, yielding an overall fidelity $\mathcal{F}_2 \approx 0.92$.
Recent progress towards the engineering of large superconducting flux-qubit arrays is particularly intriguing \cite{Johnson:2011gd,PhysRevX.4.021041};
disorder naturally arises from the fabrication process and full tomography of the couplings within the system is daunting. Thus, the ability to define MBL qubits in a modular fashion is particularly applicable. With recent coherence times \cite{PhysRevLett.113.123601} $> 10\mu$s, typical interaction strengths $\sim 1$GHz, and individual flux-qubit control, one finds a fidelity $\mathcal{F}_2 \approx 0.99$. 
In the case of molecules and solid-state spin impurities, the interactions are long-range and the dominant disorder  arises from random bonds. For effectively short-range power-laws, many-body localization persists and the main idea of this work is still applicable \cite{burin2006energy,burin2015many,Pino:2014bn,Yao:2013vc}. 

In summary, we have introduced a scheme for the coherent control of local degrees of freedom in the many-body localized phase. Our approach enables encoding quantum information as well as to perform quantum logic between separated MBL qubits. This suggests that in certain cases, strongly disordered, interacting systems  may
be a resource for quantum information applications. 
The ability to efficiently prepare (high temperature) many-body eigenstates via local spectroscopy also opens the door to studying coherent dynamics in the MBL phase. By probing the decay of both classical and quantum information, it may be possible to characterize many-body localized states and their dynamics.

It is a pleasure to gratefully acknowledge the insights of and discussions with M. C. Banuls, A. Chandran, J. I. Cirac, E. Demler, A. V. Gorshkov, J. Haah, D. Huse,  M. Knap,  C. Laumann, V. Oganesyan,  and A. Vishwanath. This work was supported, in part, by the Harvard-MIT CUA, the Kwanjeong Educational Fellowship, the AFOSR MURI, the ARO MURI, and the Miller Institute for Basic Research in Science. 
\bibliography{bib_QCMBL}

\begin{thebibliography}{52}
\bibitem [{com()}]{comment:ion_trap}%
  \BibitemOpen
  \href@noop {} {}\bibinfo {note} {This idea is analogous to the quantum
  information processing using trapped ions, where the interaction between
  ions(qubits) are mediated by photon(bus) modes.}\BibitemShut {Stop}%
\end{thebibliography}
\end{document}


\title{Supplemental Material for Quantum Control of Many-body Localized States}
\author{S. Choi,  N. Y. Yao, S. Gopalakrishnan, M. D. Lukin}
\maketitle
\noindent {\bf Numerical methods and details---}Here, we provide  full numerics for our protocol in both the spin-1 model in Eq.~(1) and the generic Heisenberg spin-1/2 model with random fields (Eq.~9).
The general outline of the demonstration is the following. 
First, for a given realization of a disordered Hamiltonian we numerically identify two qubit- and bus-type excitations, located at positions $L$ and $R$. Second, we compute the effective interaction strength between qubits either with or without off-resonant coupling to the bus excitations. Third, we repeat this process for various distances separating $L$ and $R$ and average the interaction strength over 500 realizations. Below, we provide the details of this process for both spin-1 and spin-1/2 models, always setting the lattice spacing $a = 1$.

\begin{figure}[bt]
\includegraphics[width=5in]{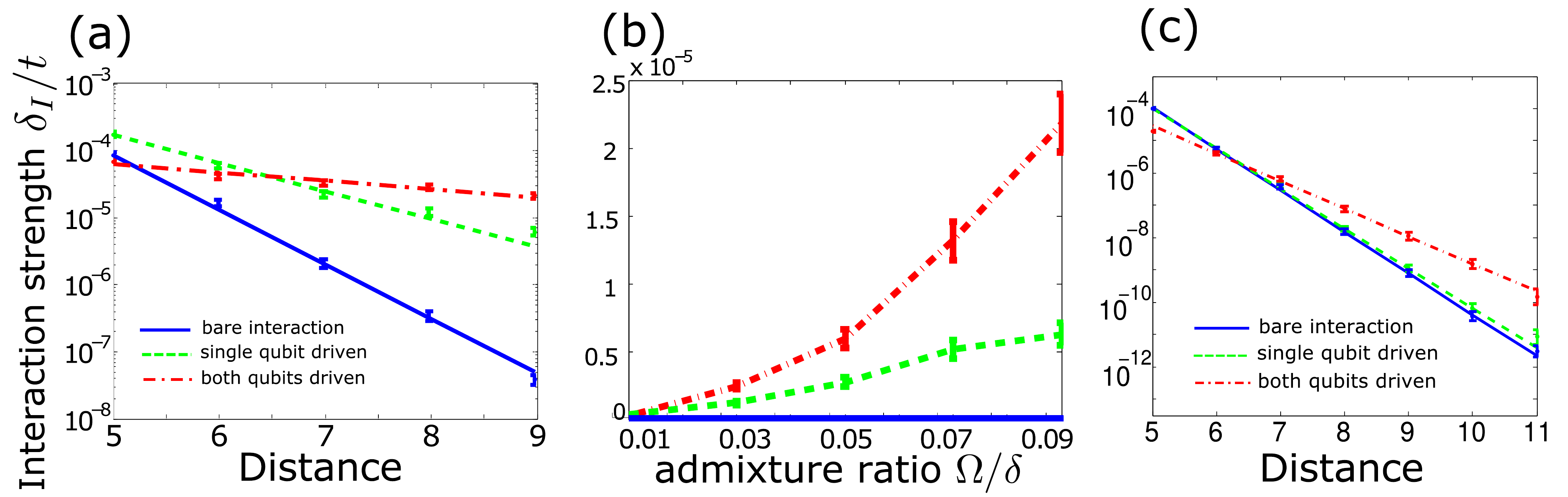}
\caption{Effective interaction strength. (a) Interaction strength decays more weakly when either one or both qubits are off-resonantly coupled to the bus excitations. (b) The enhanced interaction strength is quadratic in the ratio of Rabi frequency to detuning $\Omega/\delta$ as expected for off-resonant dressing.
(c)  Effective interaction strength decay for the random field Heisenberg model.}
\label{fig:demonstration}
\end{figure}

For the spin-1 model, we exactly diagonalize a system of 14 spins (Eq.~1) with $\Delta_0=25,\Delta = 15,h_0=h=0.02,$ and $t=1$. The Hamiltonian conserves total $z$-magnetization, $\sum_i S^z_i$. Therefore, we only need to diagonalize a few relevant symmetry sectors of the Hilbert space, namely, systems with up to four spin excitations starting from the spin polarized state $\ket{G}$.
A single qubit excited state at position $i$ is identified as the energy eigenstate with the largest overlap with $S^x_i \ket{G}$ where $S^x_i$ is the spin flip operator at position $i$. We denote the eigenstate with a single excitation at position $L$ as $\ket{\phi^L}$. In this simulation we choose $L=3$. Similarly, we identify eigenstates with single qubit excitations at various positions $R = 8$, 9, 10, 11, and 12th site and denote them as $\ket{\phi^R}$. The eigenstates with two qubit excitations $\ket{\phi^{LR}}$ and bus excitations at $L$(or $R$) are also identified in a similar way from $S^x_R \ket{\phi^L}$ and $S^x_L \ket{\phi^L}$(or $S^x_R \ket{\phi^R}$), respectively.
The off-resonant driving on the left qubit has been implemented by transforming the Hamiltonian into the rotating frame with a rotating wave approximation.
The time-dependent driving on the right qubit has been numerically integrated over one period, and then the long time dynamics is studied in the Floquet basis.
As discussed in the main text, the effective interaction strength between the MBL qubits is directly calculated from energy eigenvalues,
\begin{align}
\delta_I \equiv E_{LR} - E_L - E_R + E_G.
\end{align}
This is also the case  for the driven hamiltonian in the Floquet basis, where energies are defined via logarithms of unitary evolution.
The average interaction strength as a function of distance is shown in Fig.~\ref{fig:demonstration}a. We  confirm the exponential decay of the bare interaction strength $\delta_I$ with a length scale $\xi_1 \approx 0.53$ (blue solid line), which agrees with our simple theoretical estimate $\left(\log\frac{\Delta}{t}\right)^{-1} \approx 0.5$. When the qubits are driven, the interaction is enhanced by several orders of magnitude (red dash-dot line and green dotted line). Moreover, we observe that the enhanced interaction strength scales quadratically in $\Omega /\delta$ (see Fig.~\ref{fig:demonstration}b), confirming the off-resonant dressing picture presented in Eq.~(7) of the main text. 

We use a similar technique to simulate a system of 16 spins in the Heisenberg model, with $t = 1$ and $\Delta = 7.5 $. Again, the simulation supports up to four excitations from the polarized state $\ket{G}$.
Unlike our spin-1 model, the qubit excitations are identified by their energy eigenvalues; single spin excitations with the highest and the lowest energies are used as $\ket{\phi^L}$ and $\ket{\phi^R}$ depending on their relative positions.
A single excitation with either high or low energy is expected to have a short localization length.
Therefore, coupling to any subsequent excitations will, on average, enhance the effective interaction.
Note that our criteria of identifying qubit- and bus-excitations can be further optimized to achieve a stronger interaction enhancement.
We estimate the positions of qubit excitations $p_{L(R)}$ by maximizing $|\bra{\phi^{L(R)}} S^x_{p_{L(R)}}\ket{G}|$, and then identify bus excitations from $S^x_{p_L+1} \ket{\phi^L}$ and $S^x_{p_R-1} \ket{\phi^R}$, i.e. nearest subsequent excitations.
The effective interaction strength as a function of distance is shown in Fig.~\ref{fig:demonstration}c. Again, we confirm that the enhancement is exponentially large at long distances. We emphasize  that this model does not have `engineered' localization lengths as in the spin-1 model.

The typical localization lengths of qubit and bus excitations can be obtained from the slopes in Fig.~\ref{fig:demonstration}c, yielding $\xi_1 \sim 0.34$ (blue solid line) and $\xi_2 \sim 0.50$ (red dash-dot line). These values are consistent with the probability distribution of localization lengths studied in the next section. 

\vspace{7mm}

\noindent {\bf Variation of localization lengths---}Here, we provide numerics  showing that a variation in localization lengths always exists in the localized phase of the random field Heisenberg  model (this variation is of course most pronounced near the transition).

\begin{figure*}[tb]
\includegraphics[width=7in]{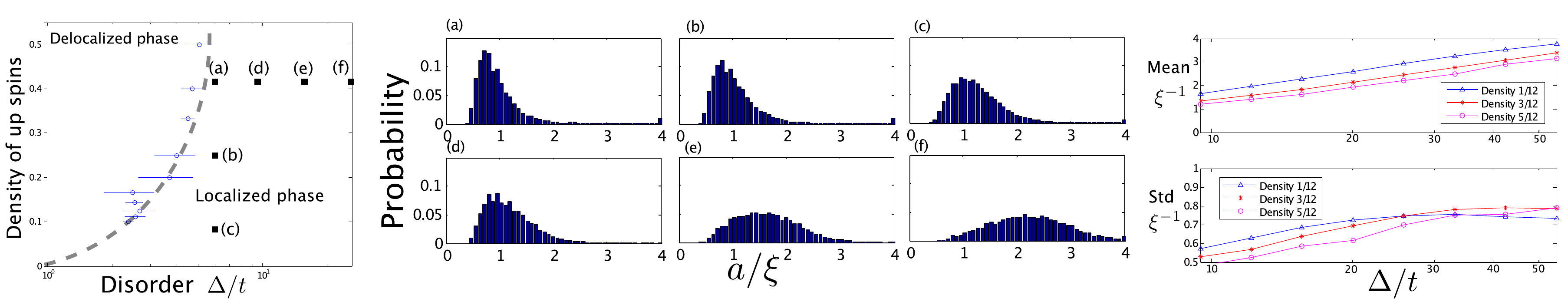}
\caption{Variation in the localization lengths. Left: the phase diagram of the random field Heisenberg spin-1/2 model as a function of spin density and disorder. Middle:  histograms of the inverse localization length for parameters shown in the left panel. Right: mean and standard deviation of the  inverse localization length in the strong disorder regime.}
\label{fig:inv_loc_hist}
\end{figure*}

First, to work in the localized phase, we estimate the criticial disorder $\Delta /t$ for the many-body localization phase transition for various spin excitation densities (relative to the polarized state $\ket{G}$).
We  follow the method used in Ref. \cite{Huse_2010}, and calculate the fraction of dynamical polarization at infinite temperature. The result is summarized in the left panel of Fig.~\ref{fig:inv_loc_hist}. We confirm that the transition occurs at $\Delta/t \sim 5$ for high densities,  consistent with previous results for small system sizes \cite{Huse_2010}.
At low densities, the critical disorder is smaller, approaching the limit of the Anderson localization criteria in 1D $\Delta / t \rightarrow 0$.

We study the variation of localization lengths under various conditions. 
Here, we \textit{define} the localization length of excitations with respect to a particular a many-body eigenstate  in the following way. First, we exactly diagonalize a system of 12 spins, choosing an arbitrary energy eigenstate with total spin up density $n_s = \sum_i \langle \left(S^z_i + \frac{1}{2} \right) \rangle /N$ and define it as our logical ground state $\ket{\mathcal{G}} = \ket{00}$. Then, we choose two ``probing positions'' $L_p = 2$ or $4$ and $R_p = 9$ or $11$ and identify all possible computational basis states (\textit{viz.} $\ket{01}$, $\ket{10}$, and $\ket{11}$) as energy eigenstates that have large enough overlap with $S^+_R \ket{00}$, $S^+_L \ket{00}$, and $S^+_L S^+_R\ket{00}$; we also impose  that these states satisfy the consistency condition $| \bra{11} S^+_L \ket{01}|, |\bra{11} S^+_R \ket{10}| > c$ with a threshold $c=0.01$.
The relative differences in the energy eigenvalues of these states \textit{defines} the interaction strength $\delta_\textrm{int}$ and the localization length $\xi$ by 
\begin{align}
d/\xi \equiv  -\log{\left( \frac{|\delta_\textrm{int} |}{t}\right)}= - \log{\left( \frac{|E_{11}-E_{10}-E_{01}+E_{00}|}{t}\right)}
\end{align}
where $E_{ab}$ is the energy eigenvalue for $\ket{ab}$ and the distance between two excitations $d$ is computed in the same way as in the previous section.
We repeat this process for different logical ground states and  disorder realizations;  various values of $n_s$ and $\Delta/t$  are shown as black squares (a-f) in  Fig.~\ref{fig:inv_loc_hist}(left). 
For each case, the histogram of inverse localization lengths  is shown in the middle panel of Fig.~\ref{fig:inv_loc_hist}. Note that we have chosen to histogram  $1/\xi$ rather than $\xi$ because the Jacobian induces long  tails in $\xi$.

The mean of the inverse localization length increases with both decreasing spin density and increasing disorder strength. 
It may seem that the variance of the distribution is also increasing with  disorder strength in (a), (d), (e), and (f); however, this behavior does not continue at larger disorder strengths. Rather, as shown in the right panel of Fig.~\ref{fig:inv_loc_hist},  the standard deviation saturates near $\Delta / t \sim 30$ while the mean continues to increase. 

\vspace{7mm}

\noindent{\bf Initialization and DEER sequences---}
Here, we provide the detailed protocol for initializing the system into an effective many-body eigenstate using a combination of DEER (double electron electron resonance) spin echo and efficient quantum  phase estimation  \cite{Knap2014, PhysRevA.74.032316}.
We assume that the quantum state of the system is initially in an unknown superposition of many-body eigenstates.
Our goal is to efficiently prepare a quantum state such that 
the coherence of qubit excitations and their interactions is maintained for sufficiently long times.
In doing so, we only use local operators and projective measurements limited to the accessible region of size $\ell$.
We assume that we can  start by preparing the small local regions $L(R)$ into their respective local eigenstates $\ket{0_{L(R)}}$ (e.g. eigenstates of $\tau^z$ spin operators within the region) \cite{Huse_2013}.

\begin{figure}[bth]
\includegraphics[width=4in]{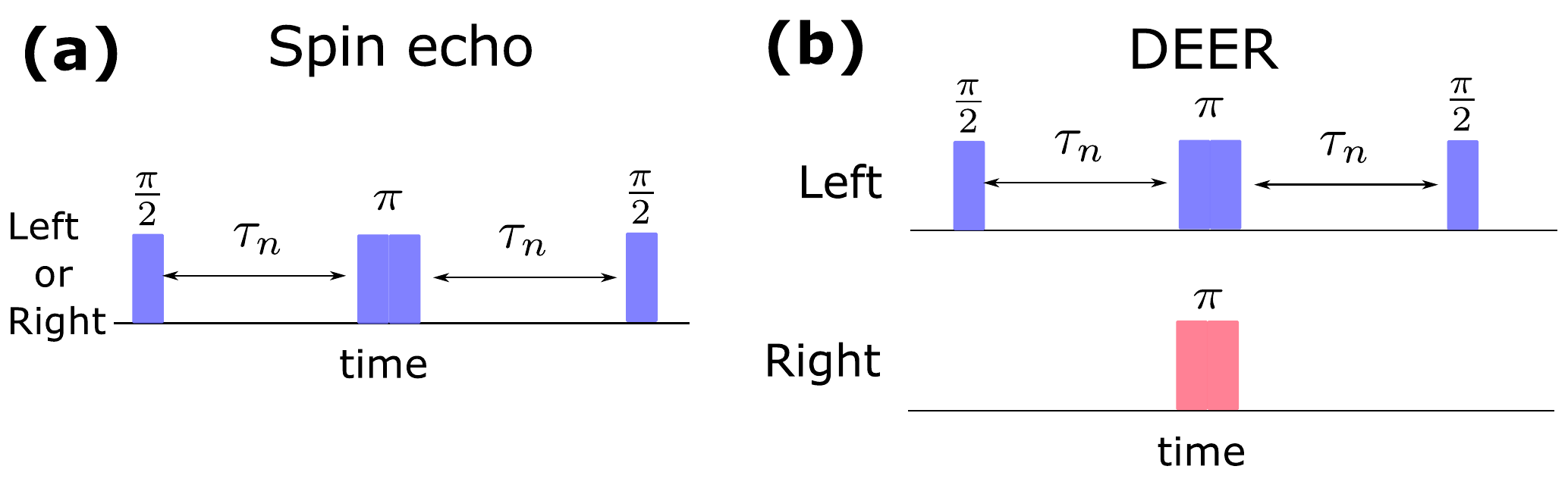}
\caption{Spin echo pulse sequences. (a) The dephasing of the left MBL qubit can be refocused via a simple Hahn echo. (b) a modified spin echo sequence (DEER) isolates the dephasing of the left MBL qubit induced by  interactions $\delta_I = \delta_{LR}+\delta_{LMR}$.}
\label{fig:pulse}
\end{figure}

The state of the full system can be written in the $\tau^z$  basis as
\begin{align}
\label{eqn:initial_step_state}
\ket{\psi_0} = \sum_M c_M \ket{0_L}\otimes \ket{M} \otimes \ket{0_R}
\end{align}
where $M$ enumerates all (exponentially many) possible configurations for the uninitialized regions. 
Then, we perform local spectroscopy in each region ($L$ and $R$) in order to estimate the energy of a single excitation at the center of the region. Due to the diagonal interactions in the MBL phase, the spectral lines for these excitations are broadened by at most $ \sim t \exp{(-l/\xi_2)} $. With $\ell \gg \xi_2$, one can  resolve and manipulate individual localized excitations in each region within a  timescale $\sim t^{-1}$. We choose a single excitation in each region to use  as our MBL qubits. These qubit excitations interact with spins in the rest of the chain and therefore dephase over a time scale $T \sim t^{-1} \exp{(\ell/\xi_2)}$.
This dephasing can in principle be refocused by the spin echo sequence considered in Fig.~\ref{fig:pulse}a \cite{Knap2014}. 

For coherent interactions between MBL qubits,  one needs to  further initialize  the  system, since the interaction strength $\delta_I$ depends on the configuration  $M$ of the uninitialized region.
For a specific $M$, the interaction can be written as $\delta_I = \delta_{LR}+\delta_{LMR}$~\cite{Huse_2013}. Here, one can understand $\delta_{LR}$ as the direct interaction between the two excitations while $\delta_{LMR}$ is the contribution provided by all of the multi-body interactions mediated by $\tau$-spins in the uninitialized region.
It is this dependence of $\delta_I$ on $M$ that that prevents the coherent interaction between MBL qubits \cite{Knap2014}.
Interestingly, this problem can be  solved by precisely measuring $\delta_I$ (to within $\delta \omega$) since the measurement serves as a quantum non-demolition measurement for the configuration of the uninitialized region.
Note that this simple solution is  only possible due to the lack of thermalization in  the MBL phase.

For the measurement of $\delta_I$, we consider a modified spin echo sequence, known as DEER, as depicted in Fig.~\ref{fig:pulse}b \cite{Knap2014}.
This sequence refocuses the dephasing of the left MBL qubit induced by all but the MBL qubit on the right \cite{Knap2014}.
Each measurement of $\delta_I$ projects the many-body state of the system into a superposition of a few eigenstates that share the values of $\delta_I$ consistent with the measurement outcomes.
Note that we do \textit{not} require the preparation of a unique eigenstate as long as the prepared state enables coherent interactions between MBL qubits up to a desired time scale $T \sim \delta \omega^{-1}$.
Repeating the DEER sequence (followed by measurements of the left MBL qubit) with a free evolution time $\tau_n$ measures the phase shift $\delta_I \tau_n$  modulo $2\pi$. Therefore, by varying the free evolution time  $\tau_n = 2^{-n}/\delta \omega$ with $n = 0, 1, 2, \dots $, one can efficiently measure $\delta_I$ up to the desired precision $\delta \omega$ \cite{PhysRevA.74.032316}. This particular sequence of $\tau_n$ is chosen such that the protocol minimizes the total evolution time and is well described in Ref. \cite{PhysRevA.74.032316}. The total measurement time $\sum \tau_n$ scales linearly with the inverse of the desired resolution $T_\textrm{init} = \eta \delta \omega$, where a constant of order unity $\eta > 2$ accounts for the repetition of the $n=0$ step due to projection noise.
  
\vspace{10mm}

\noindent{\bf Fidelity estimates---}
Here ,we analyze the fidelity of single qubit and gates under various conditions. The single qubit fidelity depends on both (i) whether the broadened spectral lines have power-law tails (e.g., a Lorentzian) or fall off more rapidly (e.g., a Gaussian) and (ii) how the spatial resolution $\ell$ compares with $\xi_1$. When $\ell \alt \xi_1$, local level repulsion ensures that there are no levels within the local spectral gap $\mathcal{E}$ of the targeted level that couple to the external driving field; when $\ell > \xi_1$, approximately $\delta\omega/\mathcal{E}$ levels per localization length are within the spatial extent of the driving, where $\delta \omega \sim 1/T_2$ is the spectral resolution. Moreover, when the linewidth is Lorentzian, the dominant imperfections are due to nearby transitions that use the tails $\sim (\delta \omega / \mathcal{E})^2$ of the Lorentzian (spectral imperfection); whereas, for rapidly decaying tails, the dominant imperfections are due to distant, exponentially weak lines that are within $\delta \omega$ of the targeted line (spatial imperfection). 
Fidelities in the four cases are summarized in Table~\ref{cases}. 

\begin{table}
\begin{tabular}{| c | c | c |}
\hline
$1 - \mathcal{F}_{1}$ & Gaussian & Lorentzian \\[3pt]
\hline
$\ell < \xi_1$ & $\exp[-2\mathcal{E} / \delta\omega]$ & $  (\delta\omega/\mathcal{E})^2 $ \\[3pt]
$\ell > \xi_1$ & $(\ell/\xi_1) \times (\delta \omega / \mathcal{E})$ & $(\ell/\xi_1) \times (\delta \omega / \mathcal{E})$ \\[3pt]
\hline
\end{tabular}
\caption{Fidelity of one-qubit gates for finite spectral and spatial resolution under various conditions.
Two different types of spectral profiles arise from different origins: Gaussian from inhomogeneous broadening, including the effects of the bare un-enhanced interaction, and Lorentzian from homogeneous broadening, e.g., extrinsic decoherence processes.}\label{cases}
\end{table}

We now focus on the fidelity of two qubit gates and derive Eq.~(12) in the main text, which we present here again for convenience,
\begin{equation}
 \mathcal{F}_2 \simeq 1- (t/\Gamma)^{-1+\xi_1 /\xi_2^{eff}}
\end{equation}
where $\Gamma$ is an effective decoherence rate and $\xi_2^{eff}$ is an effective localization length for bus excitations.

In general, the fidelity of a two qubit gate is given as
\begin{align}
1-\mathcal{F}_2 \simeq \frac{\max{\left( \Gamma, \delta_I\right)} }{\delta_I^\textrm{driven}}
\end{align}
where the numerator is the rate of decoherence due to either external imperfections or bare dephasing interactions, while the denominator is the strength of the enhanced interaction. One can always optimize $d$ such that $\delta_I (d_\textrm{opt}) = \Gamma$. Then, using $\delta_I^\textrm{driven} \sim t e^{-d_\textrm{opt} /\xi_2^\textrm{eff}}$, one can  obtain Eq.~(12). The dominant contributions to  $\Gamma$ and $\xi_2^\textrm{eff}$ vary under differing conditions. 
First, $\xi_2^\textrm{eff}$ is either $\xi_2$ or $\xi' \approx a \sqrt{tT_2}$ depending on the spectral resolution. When $1/T_2$ is larger than the typical energy spacing of bus-type excitations $\mathcal{E}_2$, one necessarilly couples a single MBL qubit with multiple excitations. In this case, the spatial extent of the qubit is broadened in the same way as that of a bound state coupled to a multi-particle contiuum of effective mass $m^* \approx 1/ta^2$, leading to $\xi' \approx \sqrt{T_2 / m^*} \approx a \sqrt{tT_2}$ \cite{Goban:1eq,PhysRevB.43.12772,PhysRevLett.64.2418}.
The effective decoherence rate $\Gamma$ is limited by the rate of population loss as well as the extrinsic decoherence scale, $\Gamma = \max{ \left( \Gamma_\textrm{loss}, 1/T_2 \right) }$, where $\Gamma_{loss}$ can be estimated using Fermi's golden rule
\beq
\Gamma_{loss} \sim \frac{1/T_2}{1/T_2^2 + \Delta^2} \left(\frac{l}{\xi_2} \Omega \right)^2 \frac{\Delta}{\mathcal{E}_2}.
\eeq
The first factor comes from the Lorentzian spectrum, the second from the effective coupling strength, and the last from the density of bus excitation states. In order to maximize  the size of the dressed qubit, $\Delta$ needs to be as small as possible,  limited only by $1/T_2$. Therefore, we set $\Gamma \sim 1/(T_2^2 \mathcal{E}_2) \left( l / \xi_2 \right)^2$, wherein the corresponding optimal spacing is given by $d_\textrm{opt} = \xi_1 \ln{\left[ t T_2 \min{ \left(1, T_2 \mathcal{E}_2 \xi_2^2 / l^2  \right) }\right]}$. 
There is a cross over at $l^2/T_2 = \mathcal{E}_2 \xi_2^2$; when $l^2/T_2 > \mathcal{E}_2 \xi_2^2$ the loss error limits the fidelity, and  $d_\textrm{opt}$ has to be relatively small in order to to avoid  long gate operation times. For $l^2/T_2 < \mathcal{E}_2 \xi_2^2$, the spectral lines of the bus-excitations are not resolved, but  the fidelity remains limited by the number of  gate operations within a decoherence window. Thus, one finds
\beq
\label{intermediatelimit}
1-\mathcal{F}_2 \simeq (tT_2)^{-1+\xi_1 /\xi'}.
\eeq
We can estimate the fundamental limit of our protocol by considering a system optimized for our purposes viz. a system that supports different types of excitations with $\xi_2 /\xi_1 \rightarrow \infty$ with perfect spatial resolution and finite decoherence time $T_2 < \infty$.  
In this case the optimal fidelity is achieved from Eq.~\eqref{intermediatelimit}.
Consequently, in the limit $\xi_2 \rightarrow \infty$ and $\xi_1 \rightarrow a$, we obtain $\mathcal{F}_2^{opt} \simeq 1 - (t T_2)^{-1 + \frac{1}{\sqrt{t T_2}}}$ as presented in the main text. 

\bibliography{bib_QCMBL}